\documentclass[alpha-refs]{wiley-article}

\usepackage{siunitx}
\usepackage{color}
\usepackage{url}
\usepackage{hyperref}
\usepackage{verbatim}

\newcommand{\tbd}[1]{\textcolor{blue}{#1}}

\makeatletter
\newcommand\footnoteref[1]{\protected@xdef\@thefnmark{\ref{#1}}\@footnotemark}
\makeatother

\papertype{}

\title{Blockchain in the Government Technology Fabric}

\author[1]{Anwitaman DATTA}

\affil[1]{School of Computer Science and Engineering, NTU Singapore}

\corraddress{Anwitaman DATTA, School of Computer Science and Engineering, NTU Singapore}
\corremail{anwitaman@ntu.edu.sg}

\runningauthor{Anwitaman DATTA}

\begin{document}

\maketitle

\begin{abstract}
Fuelled by the success (and hype) around cryptocurrencies, distributed ledger technologies (DLT), particularly blockchains, have gained a lot of attention from a wide spectrum of audience who perceive blockchains as a key to carry out business processes that have hitherto been cumbersome in a cost and time effective manner. Governments across the globe have responded to this promising but nascent technology differently - from being apathetic or adopting a wait-and-watch approach: letting the systems shape themselves, to creating regulatory sandboxes and sponsoring capacity building, or in some instances (arguably) over-regulating and attempting to put the blockchain genie back in the bottle. Possible government role spans across a spectrum: regulating crypto-currencies and initial coin offerings (ICO), formulating regulatory frameworks for managing the adoption of blockchains, particularly in critical infrastructure industries, facilitating capacity building, and finally, embracing blockchain technology in conducting the activities of the government itself - be it internally, or in using them to deliver public services. In this paper we survey the last, namely, the use of blockchain and associated distributed ledger technologies in the government technology (GovTech) stack, and discuss the merits and concerns associated with the existing initiatives and approaches. 

\end{abstract}
 
\section{Introduction}

Since around the turn of the century, with the rise of Internet penetration (around the dot.com boom and bust), many governments across the globe embraced the use of online services (\cite{heeks2001reinventing}), both as a channel for delivering services to its customers, as well as for managing the backend and internal processes. From those early days of e-government (e-gov), which the OECD defines as `\emph{the use by the governments of information and communication technologies (ICTs), and particularly the Internet, as a tool to achieve better government}'
 (\cite{OECD1}), over the next roughly two decades, we have arrived at a juncture, where rise in digital literacy, near saturation of internet penetration in advanced societies, particularly enabled by ubiquity of smartphones and mobile internet infrastructure, as well as advnaces in big data management and analytics enabled artificial intelligence (AI) techologies, put us in a sweet spot, where much more than the original objectives of e-government and e-governance (for instance, as outlined in OECD's e-government imperative document (\cite{OECDimperative})) can be achieved. The OECD emphasizes this broader scope with a distinct nomenclature `digital governance', and defines it as the ``use of digital technologies, as an integrated part of governments’ modernisation strategies, to create public value. It relies on a digital government ecosystem comprised of government actors, non-governmental organisations, businesses, citizens’ associations and individuals which supports the production of and access to data, services and content through interactions with the government'' (\cite{OECD1}).  

While often the terms e-government/governance\footnote{The distinct terms e-government and e-governance are sometimes used to distinguish the use of technology for managing the government's internal activities versus service delivery to citizens. Nevertheless, for brevity, in this paper, we will use the term e-goverment, and likewise, digital government, to capture both meanings.} and digital government are also used interchangably, the distinct OECD definitions help capture subtle yet fundamental changes that have emerged over time. Prominently, while the core mission of catering to the public at large remains the same, the means has expanded in its scope, and the emphasis has shifted from the governments delivering it on their own to creating an environment, where it can be done using public-private partnerships, as well as by facilitating purely privately funded efforts to flourish. This is being achieved by creating an ecosystem comprising digital infrastructure and regulatory frameworks on which the diverse participants can build upon. 

\begin{figure}
	\includegraphics[width=\linewidth]{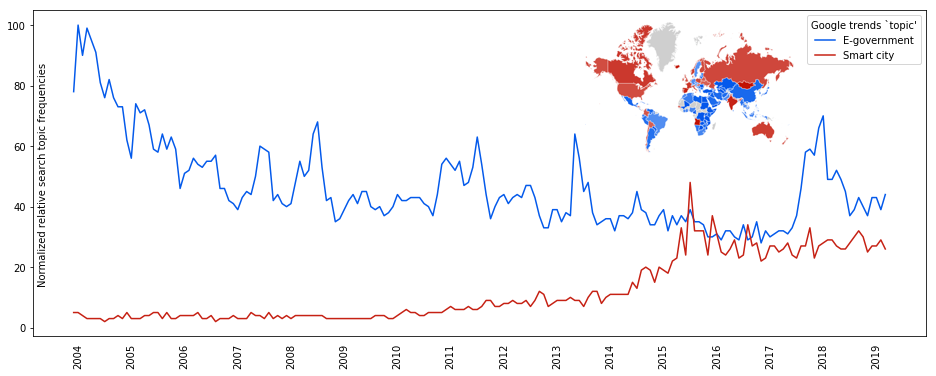}
	\caption{Data from google trends for the topics of `E-government' and `Smart city' worldwide for the period from January 2004 till April 2019 were obtained (on 2nd May 2019). The normalized relative frequencies are plotted here. The relative dominance of the individual terms in each country is also depicted in the in-set map (with the color corresponding the dominant term), obtained from the Google trends website \url{https://trends.google.com/}. For the creation of the map, low search volume regions were included (grey regions indicate lack of adequate data still).}
	\label{fig:googletrendssmartcityegov}
\end{figure}

For instance, many countries and cities have embarked on smart city initiatives in the last decade. The rise of app based ride sharing and hailing services is an example instance of how private parties are addressing urban transportation needs. Sharing economy in general, as a phenomenon (\cite{heinrichs2013sharing,martin2016sharing}), for better or worse, exemplifies this model further. In Figure \ref{fig:googletrendssmartcityegov} we show the Google topics search trend, (not to be treated as an irrefutable evidence of the argument we have put forward, but more) as one plausible indicator. We notice that searches for the topic `Smart city' have increased over time. Furthermore, the in-set map indicates that this term dominates in many of the digitally advanced countries which embraced e-government early on, and hence have moved up the data value chain. While the drive towards smart cities is just one among many aspects of digital government, it encapsulates the general trend of moving up in the data value chain, that is evolving from the original digitization drive which was focussed on creating an electronic channel for delivering citizen services and managing backend government workflows.

There are several intertwined and cascading factors at play in this evolution. What started as a move from paper based workflows to digitization, led to creation of a huge volume of data that is readily available for automated processing. The infrastructure to store and process humongous volume of data started to mature, even as the volume of data being acquired also keeps rising by leaps and bounds. This data comes from a plethora of sources, and is very diverse in nature - social media, sensors deployed for monitoring the environment, cities, buildings, financial records, health records, to name just a few. Analysing this data yields intelligence, and creates opportunities, both for solving (and identifying) problems, as well as the positive societal and financial impact such solutions yield. 

A common underlying theme in all this is the availability and sharing of (good quality of) data. However, there are several challenges that hamper this, to name a few prominent ones: data integration and portability (\cite{doan2012principles}), privacy, distributed control, provenance and data usage transparency. 


There are numerous privacy issues, and not all the issues are even well understood. This includes questions of access control (who should get what data), information leak and side channels (even if a specific data in itself may not prima facie reveal something, in conjunction with some other information, it may reveal something more than what each of the individual pieces of information disclosed). Even within different government agencies, sharing certain data may be violative of the rights of a citizen as per the laws of the country. Sharing it with non-government entities compound the concerns. Lack of (well thought of) regulations also lead to many grey areas. Furthermore, individual data aggregators need to satisfy the asscoiated privacy and security requirements, and they may also want to control the data they own in a manner where they can account for its usage.

For whichever reasons, the data in the system may be of poor quality, or outright wrong. Even otherwise, it would be reasonable to expect certain accountability on the source of the information. Thus, data provenance and lineage, along with ability to trace who all have accessed said data, and for which purposes, are also essential. 

Technological solutions are thus needed to store, process, and share data in a secure manner, balancing the needs (aspirational, as well as, often, regulatory) of utility and privacy in a highly distributed environment, involving many autonomous entities, which may not (fully) trust each other. 
Blockchain technologies have emerged as a potential candidate providing a framework to address several of these concerns. At this juncture, it is worth emphasizing that, blockchains (i) may not be the only way, or the best way, to solve the above mentioned problems, (ii) may not even be solving all the problems enumerated here (let alone other issues not mentioned). Nevertheless, it is a candidate solution that can  naturally address issues such as distributed control among untrusted entities, and provides certain extent of flexibilities, which is why they are being tried out in a plethora of application domains.

It is in this background, prompted both by need and potential but also hype, that several governments have started pilot projects of using blockchains as part of the government technology stack.

In the rest of this paper, we first briefly review distributed ledger and blockchain technologies in Section \ref{sec:background}, then in Section \ref{sec:govchains} we report (in a non-exhaustive, but representative manner) on several government blockchain initiatives, and finally in Section \ref{sec:critique} we draw our conclusions with a critique of these early efforts. 

\section{Background: Blockchains}
\label{sec:background}

\tbd{DLT versus blockchains}

National Institute of Standards and Technology (NIST, U.S. Department of Commerce) defines blockchains as ``tamper evident and tamper resistant digital ledgers implemented in a distributed fashion (i.e., without a central repository) and usually without a central authority (i.e., a bank, company, or government)'' (\cite{NIST}). Thus, one can characterize a blockchain with four salient properties: (i) immutability (append-only) data structure, (ii) replication of the data across multiple participants, (iii) whether and how participation is restricted, and (iv) how the participants establish an agreement (consensus) on how to add new data to the existing data. 

This immutability property - where any changes to existing records can be detected (tamper evident), and it is prohibitively costly if not impossible for one, or a minority of errant participants maintaining the distributed ledger to introduce illegitimate changes (tamper resistant), make blockchains particularly appealing for a wide range of applications and scenarios. Storing records of financial transactions is a natural such application, and blockchains have been popularized by their usage foremost in Bitcoin, and subsequently by a plethora of other digital (crypto)currencies which leverage on cryptography for both guaranteeing integrity of the records (as is achieved using hashing and chaining to realize a blockchain), as well as for making it prohibitive to alter past records or insert arbitrary new records (e.g., Bitcoin uses cryptographic puzzles for proof-of-work to allow addition of new blocks in the blockchain) and for the purpose of authentication and authorization (e.g., public key cryptography, specifically digital signatures are used in Bitcoin to determine the legitimacy of specific Bitcoins being spent). But many other applications, particularly whenever any form of audit is desirable, would benefit from such an immutable data structure. Even for applications, where the information being stored is inherently mutable, for example, property or land ownership records, medical history records, etc., the overlying application can leverage an immutable data-structure underneath to record the history of changes, while using the latest version as the working data. 

The replication of information across multiple sites is a necessity for enforcing the immutability and integrity of the information stored in the blockchain, where, the replicated sites act as witnesses, so that unilateral changes to past records or introduction of new records are not feasible, and instead the decisions are made by establishing a consensus on the shared state.

Since the state of the blockchain can be updated only through consensus among the multiple sites maintaining it, the blockchain can also be used to establish trust among inherently untrusted entities. Users can establish a smart contract using some programme logic (and data): for example, change the name against which a property is registered, provided a certain amount of money is paid to the original owner of said property, and rely on the blockchain to carry out the payment with tokenized money and enforce the name change.

Finally, since the information stored in the blockchain is visible to every partipating replication site in the blockchain (and whoever else have access to the date therein, more on this below), which is desirable for audit, it may appear undesirable from the perspective of privacy or confidentiality of information. However, creative workaround to preserve privacy and confidentiality may be possible for a wide range of applications, by storing a proxy information (tokenization) which in itself does not reveal the content, but carries enough information for the specific purpose of the application. For instance, if one wants to prove that a digitial document (e.g., educational certificate) was genuinely issued by certain organization, just a cryptographic hash summary of the digitally signed document may be stored in the blockchain. When the original document is presented by the owner to any specific party scrutinizing the document, the scrutinizer can validate that the document was indeed registered with the blockchain by the legitimate entity authorised to issue said document, while, the presence of the said summary in the blockchain reveals no usable information in itself. Having said that, the issue of information leak from data stored in blockchain, particularly when coupled with side information that might be available to an adversary, is still not a thoroughly understood problem, and is a topic that would need substantial further research.

\subsection{Permissionless, Permissioned and Private Blockchains}

Blockchains can be broadly classified as being permissionless or permissioned. Permissionless blockchains are open for anyone to replicate the blockchain data and participate in the (consensus) process to update the same.

Permissioned blockchains in contrast restrict those participating in the maintenance of the blockchain. This restriction might be imposed in a centralized manner or in a decentralized manner. One example of centralized membership control would be that a government agency decides who all can participate. An example of decentralized membership control could be that existing members of an alliance of organizations collectively decide to include or revoke memberships. 

Furthermore, in a permissioned blockchain, the data in the blockchain may be available (to be accessed/read) exclusively to a certain set of entities/individuals - making it a private blockchain. 

Blockchain is supposed to provide a mechanism to derive trust \cite{economisttrustmachine} among a group of entities which do not inherently trust each other. In a permissionless blockchain, where there is no a priori trust among entities, more expensive mechanisms (e.g., solving cyptographic puzzles as proof-of-work \cite{nakamoto2008bitcoin}) are required in order to mitigate malbehaving participants. In contrast, in permissioned or private blockchains, a certain degree of trust among the participants is inherently (and implicitly) assumed, and as such, it allows flexibility in the extent of stringency required to achieve consensus. 


\subsection{Forks, Side-chains and Cross-chaining}
Different members maintaning a blockchain may try to add distinct new blocks simultaneously, which leads to what is called forking. Such divergence is eventually reconciled through a consensus mechanism to arrive at a consistent shared history. However, this provides an interesting opportunity, whereby, while the participants agree on the past history that has been established, new/different rules can be imposed for defining what is a valid new block. If the new blocks are still compatible and valid as per the older set of rules (backward compatible) its called a soft work, while, otherwise, it is called a hard fork. Such forking also allows the original community of members mainaining a blockchain to split, and create new communities, with a shared history up till a time point, and different ones from the point of forking.

It may be undesirable to put all information in the blockchain, either because it overloads the system, or because of other concerns such as compatibility, or to carry out experimentations without interfering with the content of the blockchain, or simply to keep certain information off the main chain. Such cases are accommodated by running parallel ledger(s) maintained independently, which however cross-reference(s) the primary blockchain, and may also use the primary blockchain for recording certain (but not all the) states of such parallel ledgers, which are then called side-chains. 

Finally, separate blockchains may want to interchange among themselves the stored data/information or the associated digital assets, and this leads to the notion of a network of inter-operable but independent blockchains, i.e., cross-chaining \cite{crosschain}. 

  
\section{Blockchain Use-cases for Digital Government}
\label{sec:govchains}

\textbf{Notary and registry services:} The most natural application of a government service of blockchain is a notary service. Drawing on this, registry for any real world assets can also be built using a blockchain. This is indeed one of the canonical use cases of blockchain being piloted in many places \cite{georgeder,hondurasgeorgia}. Different variations exist, government led or backed initiatives, as well as private entities mirroring the government's land registry record on a blockchain. In general, registry services of all sorts can be realized using blockchain, e.g., a registry of companies \cite{smartdubai,maltaregistry}. 

\hspace{-5.5mm}\textbf{Digital identity:} The European Union Blockchain Observatory \& Forum (\url{https://www.eublockchainforum.eu}), an European Commission initiative, identifies digital identity as a fundamental building block \cite{EU-report} that can be realized using a blockchain, to act as a digital equivalent of government issued identity documents. Such a mechanism ideally needs to have flexibility, such that only relevant aspects of the digital identity are exposed, without revealing other aspects of identity information not relevant for the given context, thus affording individuals a degree of sovereignity in managing their identity. It is envisaged that such an identity service can be leveraged to provide a wide range of other services. Some pilot cases that have tried identity services using blockchains include e-voting\footnote{\url{https://medium.com/bitrates-news/swiss-city-of-zug-successfully-completes-blockchain-based-e-voting-trial-b7b312e5cdc0}} and bike rental\footnote{\url{https://medium.com/uport/zug-residents-can-now-ride-e-bikes-using-their-uport-powered-zug-digital-ids-7ed31ac9d621}} tried in Zug, Switzerland; worker training certification and check-in/out at work sites carried out by Swiss railways \footnote{\url{https://medium.com/linum-labs/swiss-federal-railway-trials-first-digital-identity-pilot-on-ethereum-4a3cb3c6621}}, debit cards for refugees without other documentations or bank account in Finland\footnote{\url{https://www.wired.com/story/refugees-but-on-the-blockchain/}}. All the three use cases from Switzerland that are mentioned here used a public blockchain based decentralized identity service (\url{https://www.uport.me}), while the Finnish initiative was in liasion with a private enterprise, Moni. Estonia, considered a pioneer in digital identity, uses a blockchain variant called KSI (Keyless Signature Infrastructure), created by Guardtime \cite{ksi}. The European Union regulation eIDAS (electronic IDentification, Authentication and trust Services) mandates a digital identity system that interoperates across EU, however it somewhat predates the recent hype around blockchain technologies, and it does not utilize blockchain as the core infrastructure to realize it, but use of and with blockchains is under deliberation\cite{eIDASBlockchain}. The ID2020 \cite{id2020} is another public-private partnership, while Decentralized Identity Foundation \cite{DIF} is an industry alliance, both aimed at realizing blockchain enabled digital identity in a global scale. In the digital identity space, we thus see use of both permissionless as well as permissioned/ private blockchains, led by government, public-private partnerships, as well as private entities. 


\hspace{-5.5mm}\textbf{Digital certificates and records:} The ideas of a registry service, along with identity, naturally extends to a repository of certificates and records. These could be inherent attributes (intrinsic to the individual, for example: data of birth/death, biometrics, health), attributes accumulated or created over time (health records, educational credentials, wealth and associated property and financial records, will \& testament), or assigned (e.g., government ID numbers, credit risk score). These records furthermore can be grouped as per the frequency with which they change \cite{illinoisreport}. Finally, a third dimension relates to access rights and privacy/confidentiality requirements of these records, namely who have the right to access which parts of said records, and who have right to update said records. The suitability of storing the spectrum of such records over blockchain, meeting the access control, confidentiality, (system) access and change logging, and content update dynamics require exploration. Blockchain use to curb degree fraud in countries as diverse as Malaysia\footnote{\url{https://www.nst.com.my/news/nation/2018/11/429615/university-consortium-set-authenticate-degrees-using-blockchain}} and Malta\cite{maltaedu} have been proposed. Hash based integrity check and logging is used for Estonia's electronic health records \footnote{\url{https://e-estonia.com/solutions/healthcare/e-health-record/}}, and many healthcare industry use cases from drug traceability \footnote{\url{https://www.ibm.com/blogs/blockchain/2018/12/what-are-the-use-cases-for-blockchain-tech-in-healthcare/}} to insurance \cite{deloitteinsurance} are expected to improve the quality of data, service and cost to deliver the same. Again, most of these are in exploratory pilot stage at the moment, and ranges from government led efforts like the one from Estonia, to alliances of private companies such as the Synaptic Health Alliance \footnote{\url{https://www.synaptichealthalliance.com}}.


\hspace{-5.5mm}\textbf{Sovereign/fiat currency:} Given that blockchain's popularity in the past decade originates from the success of cryptocurrencies, use of blockchain to support digital sovereign fiat currencies (aside the many private sector financial technology innovation attempts) is natural. Sinagpore's project Ubin\cite{projectubin}, under stewardship of the Monetary Authority of Singapore and in collaboration with major financial institutions (the Association of Banks in Singapore (ABS)) carried out pilot studies of inter-bank transactions using digital ledger technologies, and developed  three models for decentralised inter-bank payment and settlements. While the Singapore project used the blockchain technology solely as a distributed ledger for recording the transactions using tokenized coins, ``Crypto Franc'' was proposed as a bond\footnote{\url{https://www.swissinfo.ch/eng/stable-coin_crypto-bond-catapults-swiss-franc-onto-blockchain/44512880}}, with its value pegged to the Swiss franc on a 1-to-1 basis, where permissioned nodes federating Swiss Cantons were proposed to enforce the adherence to regulatory requirements, and to maintain the ledger. The status of this proposal is ambiguous at the time point this article is being written, and in general, there seems no government level support or intent for a sovereign Swiss digital currency \footnote{\label{note:swissdontwant}\url{https://cointelegraph.com/news/state-issued-digital-currencies-the-countries-which-adopted-rejected-or-researched-the-concept}}.  
A notable but dubious (because of the political background) and apparently defunct attempt was Venezuela's Oil and Mineral backed Petro currency. Marshall island, through Sovereign Currency Act of 2018 \cite{Marshall}, introduced a new blockchain based currency called the Sovereign (‘SOV’), and ironically, plans to issue physical notes for the digital currency. A July 2018 article surveys the status of state issued digital currencies on its adoption, exploration and possible rejection\footnoteref{note:swissdontwant}. 

\textbf{A platform for data management across multiple stake-holders (and digital twinning):} The term `digital twins' has roots in the cyber-physical systems community, but it can be more broadly interpreted as \cite{digitaltwin} ``A digital replica of a living or non-living physical entity. By bridging the physical and the virtual world, data is transmitted seamlessly allowing the virtual entity to exist simultaneously with the physical entity.'' In that context, a diverse range of applications, including health monitoring devices; IoT enabled `smart' solutions across a spectrum of scale - smart home, building, city; supply chains, can all be see as projecting a digital twin. Blockchain can serve as a real-time data management framework to enable digital twinning for the whole lifecycle (from data acquisition to logging its access providing provenance, integrity as well as accountability with audit trail and transparency). 

A multi-stakeholder data sharing platform, which acts as a natural aggregator of data, and allows secure data management (including, facilitating data owners sovereignity over said data) and thus enables both digital government as well as non-governmental applications (this is termed as `Blockchain platform as a service' in \cite{EU-report}) can thus be seen as the holy grail for a blockchain in the government technolgy stack. Several governments, at city, state, country as well as supra-national levels, have deployed, started developing or expressed interest in exploring the provisioning of such multi/omni-purpose blockchains. Prominent examples include Dubai \cite{smartdubai}, State of Illinois \cite{illinoisreport}, Estonia \cite{ksi}, Switzerland \cite{swissnationalblockchain}, Australia \footnote{\url{https://www.minister.industry.gov.au/ministers/karenandrews/media-releases/advancing-australias-blockchain-industry}}, and EU through European Blockchain Partnership \footnote{\url{https://ec.europa.eu/digital-single-market/en/news/european-countries-join-blockchain-partnership}}. 


\section{Critique: State and the chain}
\label{sec:critique}


The embrace of blockchains in the government technology stack is in its nascence. Consequently, despite numerous news articles, press releases and white papers, actual technical details regarding most of the initiatives are often sparse, sometimes contradictory, or just absent from the public domain in these early days. Furthermore, design and decisions may just not yet be finalized, and so things naturally change. While we have tried our best to filter out the latest relevant and correct information, it is apt to note at this juncture that some of the points we make here may inadvertently be somewhat off the mark, or may become obsolete over time.  

Just like we see diverse forms of economic systems in general, where core services for citizens (such as health care, transportation, financial services, utilities) are provided in some instances by solely government agencies, in others, by only private entities, in yet others, in private-public partnerships, and finally, also in forms where private and government run entities both operate and compete in the market; from the examples we have discussed above, we see an echo of similar different formats in the government technology (GovTech) space in general, and for blockchains for GovTech in particular. 

For the rest of our discussions here, we focus on three aspects: (i) blockchain support for digital soverign currency, (ii) blockchain as a platform for digital identity, and (iii) the nature of the underlying blockchain infrastructure. The other specific use cases, such as registry services or repository for digital records, all in turn rely on the underlying infrastructure and the identity service. 

\textbf{Digital sovereign currency:} In \cite{EU-report}, the following argument is forwarded ``Another important building block, in our opinion, is having digital versions of national currencies on the blockchain, for example through blockchain-based central bank digital currencies (CBDCs). Making it possible for legal tender to become an integral part of blockchain transactions will make it easier to reap the benefits of new technologies like smart contracts. On a systemic level, CBDCs could bring the benefits of decentralisation to inter-bank payments and real-time gross settlement systems, among other things''. 

There are several cryptocurrency flavoured approaches to realize a digital soverign currency, Venezuela's Petro (now apparently defunct) and Marshall Islands' `the Sovereign' (which also is planned to come with physical `banknotes') come to mind. While there is some level of enthusiasm about such digital soverign currencies (to be distinguised with the non-state-backed cryptocurrencies), it is rather unnecessary and a tokenization based approach showcased in project Ubin is a pragmatic solution to realize central bank digital currencies (CBDCs).  

To quote \cite{projectubin}: ``the SGD-on-ledger is a specific use coupon that is issued on a one-to-one basis in exchange for money. The coupons have a specific usage domain – in our case for the settlement of interbank debts – but no value outside of this. One is able to cash out by exchanging the coupons back into money later ... SGD-on-ledger has three useful properties that make it suited to our prototype. First, unlike money in bank accounts, we do not receive interest on the on ledger holdings. The absence of interest calculations reduces the complexity of managing the payment system. Second, to ensure full redeem-ability of the SGD-on-ledger for money, each token is fully backed by an equivalent amount of SGD held in custody. This means that the overall money supply is unaffected by the issuance of the on ledger equivalents since there is no net increase in dollar claims on the central bank. Third, SGD-on-ledger are limited use instruments and can be designed with additional features to support the use case – such as security features against misuse.'' 

The three highlighted properties from the project Ubin report emphasize the importance of responsibly using blockchain without creating instability while solving actual pain points of digitial financial activities that exist with legacy infrastructure: particularly that the processes are unnecessarily complex with respect to the functionalities provided, making the solutions inefficient (slower and/or expensive). Such a tokenized approach also allows a natural integration of the currency with other workflows and functionalities that may be carried out over a multi/omni-purpose blockchain platform.  

\textbf{Digital ID in a multi-stakeholder environment:} Digital identity has been repeatedly emphasized as one of the `killer apps' of government blockchains. For instance, to quote \cite{illinoisreport}: ``A citizen-centric digital identity model based on distributed ledger technologies could be used to consolidate disparate data that currently exists across multiple agencies and layers of government into a network centered around a citizen’s or business’ credentials, licenses and identity attributes. It would enable citizens to view their public service identity via an identity app on their smartphone and share relevant data with government to access public services.'' A European Union Blockchain Observatory \& Forum report \cite{EU-report} likewise states ``One of the most important requirements in building a digital economy and society is viable digital identities for all participants, whether individuals, companies, public agencies or, increasingly, machines and other autonomous agents. The need to be able to identify ourselves and others is so important, in fact, that it is considered the essential prerequisite for most use cases.'' Many other whitepapers \cite{WEF-ID,WorldBank-ID,australiaID,self-sovereignID} have likewise elaborated the importance of digital identity in the recent years.

The Estonian blockchain at its core deals with digital identity \cite{ksi}, and several of the proposed government run blockchains aim to provide and utilize digital identity in some manner. Yet, some of the world's largest digitalized identity systems are in fact not blockchain based. This includes EU's electronic IDentification, Authentication and trust Services (eIDAS \footnote{\url{https://ec.europa.eu/digital-single-market/en/trust-services-and-eid}}), India's Aadhaar \footnote{\url{https://uidai.gov.in/my-aadhaar/about-your-aadhaar.html}} which is the world's largest world's largest biometric ID system managed by Unique Identification Authority of India (UIDAI) and China's social credit system \cite{economistchinesesocialcredit}. So the performance at scale, or the multi-stakeholder usage scenarios in themselves do not neccessitate the use of a blockchain. One argument for using blockchains is the notion of `self-sovereign digital identity' \cite{self-sovereignID}. In this (as well as many other security benefits that are assumed and/or promised with blockchain, such as more generally data sovereignty, portability, privacy and security, integrity and audit trail), the nuances of how the underlying infrastructure is actually designed, deployed and used determines whether the security guarantees are actually realized. It is too early to comment on how \cite{id2020} or \cite{DIF} technology stack evolves, but we will thus discuss (below) the specific model used in \cite{ksi}. Overall, what is factual is that blockchain based distributed ledger technology can be used to support digital identity. However, it is not a singular option to do so, nor are all the assumed security guarantees inherent invariants. 

\textbf{The infrastructure behind a blockchain as a multi/omni-purpose platform:} As recently as earlier this year, on 9 February 2019, the European Medicines Verification System (EMVS)\footnote{\url{https://emvo-medicines.eu/mission/emvs/}} was launched\footnote{\url{https://emvo-medicines.eu/new/wp-content/uploads/EMVO-Press-Release-EMVS-Launch.pdf}}. Such an application perfectly fits a blockchain use case, given the scale and multi-stakeholder nature of the system. It (to the best of our understanding) however does not use blockchain technologies.
In any case, many large-scale multi-stakeholder systems in general exist and operate meeting their design objectives. While technologically not singular, and many other alternative realizations are possible (as exemplified by deployed systems), one argument in favour of a blockchain is to expose it as a platform or service, where new applications can be modularly integrated, rather than having to design and deploy different systems from scratch for individual applications. Incidentally, since such systems are being built ground up, they are in a position to avoid some of the problems faced by many legacy systems, such as poorly structured, non-standardized data, interoperability across systems (though blockchain interoperability is still an open research issue \cite{EU-reportinfra,crosschain,crosschainmyth}) and the ability to migrate the data to/from another system. While these are not inherent properties of blockchain, the creation of a new digital infrastructure provides a coincidental opportunity.   

In \cite{EU-reportinfra,EU-report}, some design dilemmas are discussed at length. For instance, a top-down approach where the government deploys (and possibly enforces) the usage of a single blockchain for every government related purpose, will help with the aforementioned standardization by default, and yet, it may lead to a single vendor lock-in, while lacking the flexibility to accommodate all possible use cases. Many of the national blockchain initiatives \cite{smartdubai,swissnationalblockchain,ksi} appear to be following this approach of a single standardized blockchain. In contrast, uncoordinated experimentations of different technologies by different agencies may lead to duplication of effort, as well as fragmentation of platforms. In \cite{EU-report} a middle ground is advocated: ``flexible, cloud-based shared infrastructure that hosts different protocols as well as developer tools, and an integrated development and operations environment''. The authors further add ``A shared “sandbox” approach, even one featuring multiple technologies, should also foster knowledge sharing and make it easier for agencies to work together to ensure interoperability''. Particularly for a supra-national set-up such as the EU, this approach may be inevitable, since individual member states would likely embrace a spectrum of blockchain solutions.       

While the above design dilemmas are relevant, in this paper, we want to highlight a few other, arguably more critical issues, that needs careful attention. 

Consider the KSI blockchain (used in \cite{ksi}), quoting \cite{ksidatasheet} regarding data privacy guarantees: ``KSI does not ingest any customer data; data never leaves the customer premises. Instead the system is based on one-way cryptographic hash functions that result in hash values uniquely representing the data, but are irreversible such that one cannot start with the hash value and reconstruct the data - data privacy is guaranteed at all times.''. Since the blockchain does not store the actual data, prima facie data privacy is achieved using the blockchain, while also validating data integrity. However, depending on the nature of the data/application - if it is something that resides on an off-chain storage repository and is corrupted, the blockchain would be able to detect such corruption upon usage of said off-chain data, but it does not support prevention or correction (for which, out of chain mechanisms would be required in a well designed system). Likewise, the confidentiality of such data may still be violated if the off-chain repository is breached. For the electronic voting system i-Voting, \cite{ksi} states: ``the Estonian solution is simple, elegant and secure. During a designated pre-voting period, the voter logs onto the system using an ID-card or Mobile-ID, and casts a ballot. The voter’s identity is removed from the ballot before it reaches the National Electoral Commission for counting, thereby ensuring anonymity. With any method of remote voting, including traditional postal ballots, the possibility of votes being forced or bought is a concern. Estonia’s solution was to allow voters to log on and vote as many times as they want during the pre-voting period. Since each vote cancels the last, a voter always has the option of changing his or her vote later.'' However such broad and strong claim of security calls for skepticism. For instance, side-information such as time of authentication/communication might reveal who a specific person voted for, even if that information is not explicitly stored. We are not asserting that the Estonian blockchain deployment in their e-Government technology stack necessarily suffers from all these vulnerabilities, and in fact, it is very likely that some these concerns have been looked into and many further layers of protection have been deployed. The purpose of this discussion using hypotheticals is to emphasize that the blockchain does not and cannot provide a range of security guarantees in a stand-alone manner, yet we often see a marketing pitch in the lines of `its secure because it is a blockchain'. 

In \cite{ksi} it is further stated: ``With KSI Blockchain deployed in Estonian government networks, history cannot be rewritten by anybody and the authenticity of the electronic data can be mathematically proven. It means that no-one – not hackers, not system administrators, and not even government itself – can manipulate the data and get away with that.'' 

However, generally speaking the claim that data is immutable because it is on a blockchain (which is the one fundamental functionality a blockchain is supposed to provide) may overlook some fundamental issues. In the specific case of \cite{ksi}, the blockchain is further published in the physical media (news papers), which are subscribed by many libraries spread worldwide, creating a globally dispersed physical backup which is nigh to impossible to tamper. 

Public blockchains are hugely inefficient for the purposes of e-Government use cases. From usability and cost perspectives, the sole rational choice is to use permissioned (and even private) blockchains. For instance,  \cite{swissnationalblockchain} states ``Swiss Post and Swisscom are connecting their existing private infrastructures for blockchain applications. On the basis of distributed ledger technology, the two instances check each other and thus help to establish trust. In contrast to "public blockchains" (e.g. Bitcoin and Ethereum), this private blockchain infrastructure requires much less energy, since it can only be used by identified users who have a contractual relationship with the providers of an application. This enables more efficient agreement procedures as well as significantly higher security and performance. This is an important prerequisite for many companies to launch their own applications based on blockchain technology.'' 

The participating entities in such permissioned or private blockchains can collude together, or may be forced by a (hypothetical, dystopian) government, to manipulate the data. Furthermore, in many such deployments, the software running at all sites are sourced from the same vendor. So a software (update) run by all the sites from a malicious or compromised vendor would be sufficient to subvert the whole blockchain's integrity. These are some very critical issues that need more attention, particularly in the context of blockchain use in the governemnt technology stack. An approach like \cite{ksi} utilizing off-chain globally dispersed physical back-up is a nifty safeguard.

Finally, to wrap up, we want to note that smart contracts can be used over a `Blockchain platform as a service' to automate many tasks, including, near real time monitoring and actuation of action plans, and in the longer term, to enhance workflows and decision processes further driven by analytics (artificial intelligence). Such automation can significantly improve the cost effectiveness and quality of service that can be delivered. However, the opportunities to leverage such data using the blockchain infrastrucutre directly may also be constrained, depending on the nature of the blockchain deployment. For example, if the actual data is stored off-chain, and tokenization is used, then the nature of tokenization would influence the versatility of applications that can be built on top of the blockchain. This in itself is not necessarily a bad thing, nor does it add fundamental limitations in creating decentralized applications leveraging the troika of blockchains, smart contracts and artificial intelligence. In \cite{lopezdatta}, an argument for (a network of) blockchains being utilized as a glue to bind actual data and services that are off the chain (to realize better data soverignity), and likewise keeping the logic also at the edge (which is where the data originates and/or is utilized) is forwarded.



\bibliography{sample}

\end{document}